\documentclass[pdflatex,sn-mathphys-num]{sn-jnl}

\usepackage{graphicx}%
\usepackage{multirow}%
\usepackage{amsmath,amssymb,amsfonts}%
\usepackage{amsthm}%
\usepackage{mathrsfs}%
\usepackage[title]{appendix}%
\usepackage{xcolor}%
\usepackage{textcomp}%
\usepackage{manyfoot}%
\usepackage{booktabs}%
\usepackage{algorithm}%
\usepackage{algorithmicx}%
\usepackage{algpseudocode}%
\usepackage{listings}%
\usepackage{tabularx}
\usepackage{booktabs}
\usepackage{multirow}

\usepackage{lineno}

\theoremstyle{thmstyleone}%
\theoremstyle{thmstyletwo}%

\theoremstyle{thmstylethree}%

\begin{document}


\title{Black-Box Coherence Matrix Eigen-Spectroscopy with Programmable Photonics}


\author*[1]{\fnm{Kevin} \sur{Zelaya}}\email{kzmeme@rit.edu}

\author[1]{\fnm{Jonathan} \sur{Friedman}}

\author*[1]{\fnm{Mohammad-Ali} \sur{Miri}}\email{ali.miri@rit.edu}

\affil[1]{\orgdiv{Department of Electrical and Microelectronic Engineering}, \orgname{Rochester Institute of Technology}, \orgaddress{\street{1 Lomb Memorial Drive}, \city{Rochester}, \postcode{14623}, \state{NY}, \country{USA}}}

\abstract{
The precise characterization of spatial optical coherence is fundamental to emerging applications in optical communications and computational imaging. However, extracting the full coherence matrix traditionally requires phase-sensitive interferometry, which is highly vulnerable to environmental noise and poses severe scalability challenges for integrated photonics. Here, we introduce an architecture-agnostic framework for analyzing and controlling partially coherent light on programmable photonic circuits. By leveraging the Schur-Horn theorem, our approach systematically diagonalizes the incident coherence matrix, relying solely on output intensity measurements and entirely circumventing the need for complex phase retrieval. We experimentally validate this black-box protocol on a low-depth, non-universal photonic integrated circuit, successfully reconstructing the hidden eigenvalues of mixed states generated from up to four mutually incoherent sources. Furthermore, we demonstrate active, in-situ statistical light control by introducing non-unitary amplitude modulation to significantly enhance interference visibility, exposing a fundamental physical trade-off between coherence enhancement and optical loss. Inherently resilient to hardware constraints and experimental noise, this scalable paradigm establishes a robust pathway for realizing ultra-compact, on-chip spatial coherence analyzers.
}

\keywords{statistical light, photonic integrated circuits, unitary evolution, optical analog computing}



\maketitle

\section*{Introduction}

Optical coherence fundamentally dictates the capacity of a light field to form stable interference patterns, serving as the cornerstone for classical phenomena ranging from foundational diffraction and stellar interferometry~\cite{born2013principles} to advanced imaging modalities such as holography~\cite{gabor1948new,huang2024quantitative} and optical coherence tomography (OCT)~\cite{huang1991optical}. The capabilities of a light source are intrinsically tied to its degree of coherence. Fully coherent sources, such as conventional lasers~\cite{siegman1986lasers} and free-electron lasers (FELs)~\cite{mcneil2010xray,pellegrini2016physics}, maintain deterministic phase relationships across extensive spatiotemporal scales, setting a standard for high-contrast interferometry and holography. Conversely, fully incoherent sources, such as ideal blackbody radiators, exhibit chaotic, uncorrelated phase fluctuations. Between these extremes exists the regime of partially coherent light; even fundamentally incoherent thermal radiation gains partial spatial coherence upon propagation, a phenomenon mathematically governed by the Van Cittert-Zernike theorem~\cite{born2013principles}. Modern partially coherent sources, including superluminescent diodes (SLDs)~\cite{bohman2006optical}, synchrotron radiation~\cite{attwood2017xrays}, and laser speckle fields~\cite{goodman2007speckle}, possess finite correlation lengths. Rather than an artifact, engineering this partial coherence is increasingly recognized as a critical tool for mitigating speckle degradation in projection systems~\cite{lee2020light} and enhancing resolution in imaging modalities such as optical coherent tomography (OCT) and X-ray phase-contrast imaging.

To dynamically characterize and manipulate the spatial coherence of diverse radiation fields, programmable photonic integrated circuits (PICs) offer a versatile platform~\cite{bogaerts2020programmable}. By confining complex interference networks to on-chip platforms with reduced footprint, PICs circumvent the severe alignment sensitivities and calibration overheads inherent to free-space optics. The foundation of this on-chip control lies in programmable linear discrete unitary transformations, $U(N)$. Traditionally, universal optical manipulation is achieved by decomposing an arbitrary $U(N)$ operator into a cascaded mesh of tunable two-mode $U(2)$ operations~\cite{reck1994experimental,de2018simple}, physically realized via Mach-Zehnder interferometers (MZIs)~\cite{miller2013self} or ring resonators~\cite{yi2021multi} arranged in rectangular~\cite{clements2016optimal} or diamond-like~\cite{Shokraneh2020,rahbardar2023addressing} topologies. Recent architectures have addressed the restrictive fabrication tolerances and scaling limitations of $U(2)$ decompositions by bypassing $U(2)$ blocks and adopting $U(N)$ block-based decompositions. Indeed, by interlacing $N$-port couplers in the form of waveguide arrays~\cite{markowitz2023universal}, or multimode interferometers (MMIs)~\cite{pastor2021arbitrary} with programmable phase shifters~\cite{tanomura2022scalable,zelaya2024goldilocks,friedman2025programmable}, these architectures provide highly robust, error-resilient platforms for arbitrary unitary synthesis~\cite{Markowitz2023auto}.

Building on the advanced capability to route and synthesize optical fields within photonic networks, recent efforts have sought to establish deterministic and on-chip metrics for the degree of coherence. Notable progress has been achieved by exploiting the intrinsic routing of triangular MZI meshes, where light is guided layer-by-layer to maximize localized intensities~\cite{roques2024measuring}. Similarly, hexagonal MZI networks have leveraged their unique internal topologies to reconstruct generalized Stokes parameters and, consequently, the full coherence matrix~\cite{hashemi2026chip}. Despite these advances, current methodologies for partial coherence analysis remain rigidly tied to specific network architectures and rely heavily on precise, error-free calibration of individual hardware components.

In this work, we propose and validate a universal framework for characterizing and manipulating partially coherent light, independent of the underlying PIC topology. By leveraging the Schur-Horn theorem and treating the photonic mesh as a black-box system, we demonstrate that the eigenvalues of the input coherence matrix can be deterministically extracted relying solely on output intensity measurements. This architecture-agnostic approach offers a twofold advantage. First, it provides intrinsic resilience against fabrication imperfections and optical crosstalk, empowering an in situ optimization protocol to accurately retrieve coherence eigenvalues without requiring a perfectly calibrated mesh. Second, the reconstruction protocol remains robust even when deployed on non-universal PIC networks, paving the way for low-depth and footprint-efficient on-chip coherence analysis. Finally, by incorporating tunable lossy elements, we show that this framework naturally extends to active coherence control, provided the associated optical losses can be tolerated.

\section*{Results}
\subsubsection*{Theoretical Background}
\begin{figure*}[t!]
    \centering
    \includegraphics[width=0.85\linewidth]{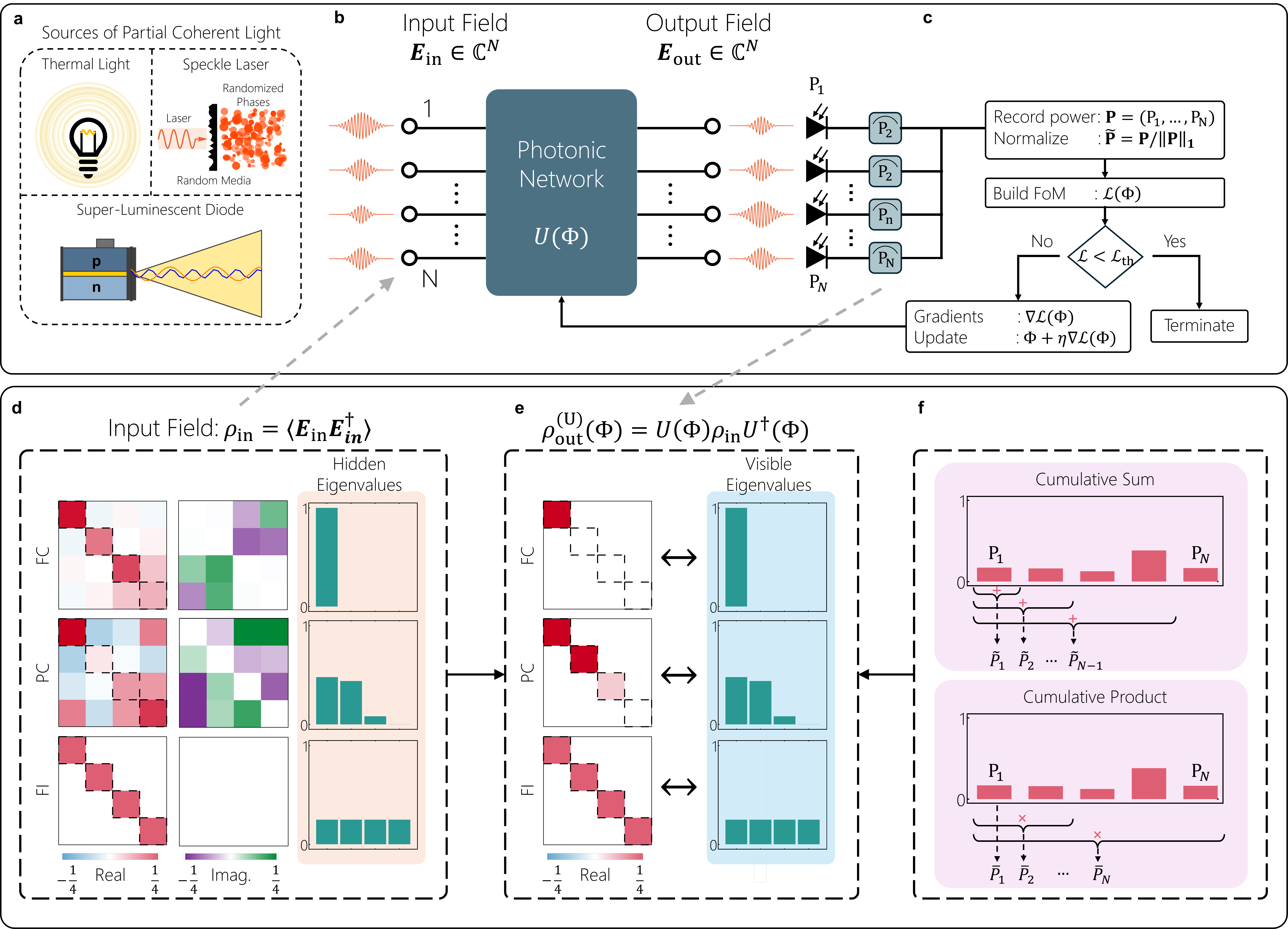}
    \caption{\textbf{Scheme for Statistical Light Analysis with Programmable Photonics.} 
    \textbf{a} Examples of statistically uncorrelated light. 
    \textbf{b} Photonic integrated circuit approach for the analysis of statistical light. This comprises a programmable, yet unknown, unitary network $U(\Phi)$ that transforms the input field $\mathbf{E}_{\textnormal{in}}$ and coherence matrix $\rho$ into the output field $\mathbf{E}_{\textnormal{out}}$ and coherence matrix $\rho^{(U)}_{\textnormal{out}}(\Phi)$, respectively. 
    \textbf{c} Architecture-agnostic approach to processing and analyzing light based solely on intensity measurements and gradient-descent optimization, without prior knowledge of the unitary network. 
    \textbf{d} Typical examples of coherence matrices and their related hidden eigenvalues for fully coherent (FC), partially coherent (PC), and fully incoherent (FI) light. Dashed rectangles highlight the available intensities, the only accessible quantity in the experimental run. 
    \textbf{e} Processed and diagonalized coherence matrices for each case in \textbf{d} using the unitary network $U(\Phi)$, highlighting how the intensities unravel the hidden eigenvalues. 
    \textbf{f} Cumulative power sums and products used for coherence matrix diagonalization.}
    \label{fig:F1}
\end{figure*}

For the present work, we focus on electromagnetic radiation fields and their statistical correlations 
of up to second order. Various relevant physical fields can be fully characterized under this scheme; sources of this kind include thermal radiation, light-emitting diodes (LEDs), speckle lasers, which are prime examples of partially coherent light, and continuous-wave lasers as fully coherent sources. See Fig.~\ref{fig:F1}\textbf{a}. Let us consider an electromagnetic field, $\mathbf{E}(\mathbf{r},t)$, propagating in space and collected at a discrete set of spatial points $\mathbf{r}_{n}$, with $n\in\{1,\ldots,N\}$. By restricting to a single polarization component, as will be the case in the subsequent discussion and experimental setup, the sampled electromagnetic field $\mathbf{x}(t)\in\mathbb{C}^{N}$ is represented by the complex-valued column vector
\begin{equation}
\mathbf{x}(t):=\left( E(\mathbf{r}_{1};t),\ldots,E(\mathbf{r}_{N};t) \right)^{T} .
\label{eq:fields1}
\end{equation}

The phases at various points fluctuate according to the nature of the source field. For fully coherent sources, the relations between phases at each sampled point can be established in a deterministic form. In turn, for fields with phase fluctuations, the field becomes incoherent, and the phases at different spatial points cannot be established directly. Thus, it is more reasonable to define a quantity that can account for such effects while keeping its measurable nature. This was indeed the reasoning behind Wolf's original work~\cite{wolf1982new}, where the spatial two-point \textit{coherence matrix} $\rho$~\cite{mandel1995optical,born2013principles} is the main object of study. The latter is defined in terms of the sampled radiation field $\mathbf{x}(t)$ via 
\begin{equation}
\rho = \langle \mathbf{x}(t) \mathbf{x}^{\dagger}(t) \rangle,
\end{equation} 
where $\langle \cdot \rangle$ is the ensemble average, which, under the assumption of statistical stationarity and ergodicity, is equivalent to the time average of the field~\cite{goodman2015statistical}.

The matrix $\rho$ explicitly encodes the spatial correlations between all pairs of sampling points, where its diagonal elements yield the average intensities at each point, and the off-diagonal elements describe the mutual coherence between distinct locations. From the definition of $\rho$, and assuming statistical stationarity, the coherence matrix is inherently Hermitian ($\rho = \rho^{\dagger}$) and positive semi-definite. These mathematical properties guarantee that $\rho$ can be factored via its eigen-decomposition into 
\begin{equation}
\rho = V\Lambda V^{\dagger}, \quad V \in \mathcal{U}(N), \quad \Lambda=\textnormal{diag}(\lambda_{1},\ldots,\lambda_{N}),
\end{equation} 
with $V$ a unitary matrix whose columns are the eigenvectors of $\rho$, and $\lambda_{n}$ the $n$-th real and non-negative eigenvalue. We assume that the eigenvalues are sorted in descending order, $\lambda_{n}\geq\lambda_{n+1}$. Physically, this decomposition serves as the discrete analog to the coherent mode representation of optical fields~\cite{wolf1982new}, where the eigenvectors represent mutually uncorrelated, fully coherent spatial modes inherent to the field, while each eigenvalue $\lambda_{n}$ quantifies the average intensity contributed by its respective mode.

\subsubsection*{Statistical Light Analysis via Photonics}
The transformation of the electric fields can be expressed in an equivalent formulation in terms of the coherence matrix, which is the main object of study. From an experimental standpoint, we can only access the power measurements of the coherence matrix to unravel its eigenvalues. In this work, we focus on $N$-dimensional coherence matrices and a discrete unitary transform $U(\Phi)\in\mathcal{U}(N)$ that steers the dynamics of the coherence matrix. Here, $\Phi$ is a set of tunable parameters that tunes the unitary transform, which in programmable PICs is typically related to phase elements~\cite{miller2013self,clements2016optimal,tang2021ten,zelaya2024goldilocks}. Since the sampled electric fields at the input $\mathbf{x}$ transforms via a matrix-vector multiplication rule, $\mathbf{x}_{\textnormal{out}}=U(\Phi)\mathbf{x}$, the associated coherence matrix $\rho$ evolves via a unitary transformation of the form
\begin{equation}
    \rho^{(U)}_{\textnormal{out}}(\Phi):= U(\Phi) \rho U^{\dagger}(\Phi) ,
\end{equation}
where the latter holds for statistically stationary fields. The exact form of $U(\Phi)$ depends on the PIC topology, which include MZIs networks~\cite {reck1994experimental,miller2013self,clements2016optimal,saygin_robust_2020} and networks involving multiport couplers~\cite {tang2021ten,zelaya2024goldilocks,friedman2025programmable}. The specifics of the design of each architecture follow from the evanescent coupling between waveguides, governed by coupled-mode theory~\cite{Huang94}.

Such measurements correspond to diagonal components of the output coherence matrix, $(\rho_{\textnormal{out}})_{nn}$. Thus, from the eigen-decomposition, if the coherence matrix is already diagonal, the power measurements correspond (up to a normalization factor) to the singular values of the output radiation field. Mathematically, this is done if the parameters $\Phi$ of the unitary network are chosen such that $U(\Phi)=V^{\dagger}$, reducing the output coherence matrix to $\rho_{\textnormal{out}}=\Lambda$, from which the analysis of the coherence degree becomes immediate. The diagonalization approach has been shown to be successful in Ref.~\cite{roques2024measuring} via a photonic triangular unitary network with the topology of Reck \textit{et al.}~\cite{reck1994experimental}. Likewise, a similar approach has been implemented in Ref.~\cite{hashemi2026chip} with a hexagonal-like photonic unit. 

In general, the exact details of the network topology are not accessible, and performance cannot be guaranteed in non-ideal scenarios. These issues can be addressed by implementing the architecture-agnostic approach in Ref.~\cite{zelaya2026analysis}, making the coherence analysis available across different network topologies. This approach is independent of the unitary photonic architecture, treats the PIC as a black-box unit (Fig.~\ref{fig:F1}\textbf{b}), and relies on the Schur-Horn theorem (see Methods section). To this end, without any prior knowledge of the PIC unitary network or the incident field, we can unravel the eigenvalues by minimizing a predefined figure of merit (FoM) $\mathcal{L}(\Phi)$ based on intensity measurements and conventional gradient-descent optimization routines, as illustrated in Fig.~\ref{fig:F1}\textbf{c}. In particular, we define the following two FoMs: 
\begin{equation}
    \underset{\Phi\in\mathcal{S}}{\textnormal{min}}\mathcal{L}^{(j)}(\Phi) ,\quad \mathcal{L}^{(j)}(\Phi) = \frac{\Vert \widetilde{\mathbf{P}}^{(j)}(\Phi) \Vert^2}{N}, \quad j\in\{1,2\} , 
    \label{eq:loss1}
\end{equation}
where $\Vert\mathbf{z}\Vert$ the Euclidean norm of $\mathbf{z}\in\mathbb{R}^{N}$, the real-valued vectors $\widetilde{\mathbf{P}}^{(1)}(\Phi)\in\mathbb{R}^{N-1}$, and $\widetilde{\mathbf{P}}^{(2)}(\Phi)\in\mathbb{R}^{N}$ define the cumumlative sum and cumulative product for $j=1$ and $j=2$, respectively, defined as
\begin{equation}
    \widetilde{\mathbf{P}}^{(1)}_{n}= 1-\sum_{k=1}^{n}P_{k}(\Phi), \quad
    \widetilde{\mathbf{P}}^{(2)}_{m}= 1-P_{m-1}(\Phi)P_{m}(\Phi), \quad 
    \label{eq:loss2}
\end{equation}
for $n\in{1,\ldots,N-1}$ and $m\in{1,\ldots,N}$, together with $P_{-1}(\Phi)=1$ and $P_{n}(\Phi)=(\rho^{(U)}_{\textnormal{out}}(\Phi))_{nn}$ the normalized measured intensities. The original goal is to maximize the cumulative sum and products in both FoMs; however, we have converted the problem into a minimization one by subtracting the cumulative quantities from one. Since the intensity measurements are normalized, the FoMs are always positive semi-definite. Still, the optimal solution is not found when the FoMs reach zero, and the minimal value is unknown as it depends on the nature of the statistical light fed at the input. Indeed, $\mathcal{L}^{(1)}(\Phi)$ converges to zero only if the input is a fully coherent source. 
For a thorough numerical testing using the cumulative sum algorithm, see Ref.~\cite{zelaya2026analysis}.

The simple yet robust approach presented here enables an error-resilient implementation, for if individual components deviate from ideal behavior due to thermal crosstalk or other unwanted effects, our approach treats the entire device as a black box and compensates the active components as needed. In this form, the hidden characteristics of the input sources (Fig.~\ref{fig:F1}\textbf{d}) can be analyzed using Eqs.~\eqref{eq:loss1}-\eqref{eq:loss2}. Minimizing such equations ultimately yields the optimal currents that diagonalize the input coherence matrices. After normalization, the intensities $P_{n}(\Phi)$ at the network output reveal the hidden eigenvalues in descending order (Fig.~\ref{fig:F1}\textbf{e}), enabling the classification of the statistical nature of the input radiation fields. Furthermore, since the unitary network $U(\Phi)$ is programmable, it can be configured to yield the identity matrix at the end of the analysis, allowing the light to be routed for subsequent use.

\subsubsection*{Fabricated Photonic Network}

\begin{figure}
    \centering
    \includegraphics[width=0.95\linewidth]{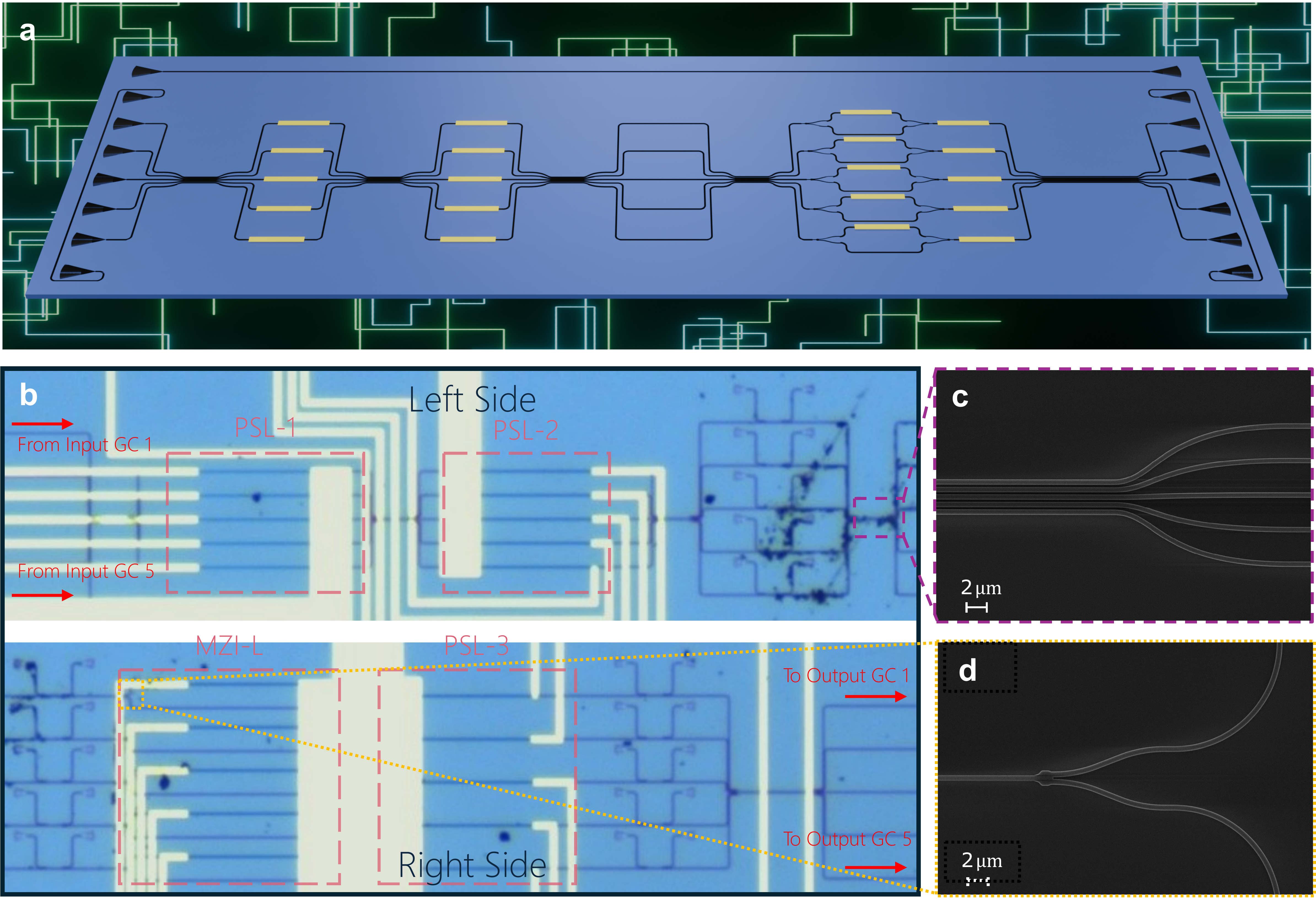}
    \caption{\textbf{Photonic network and fabricated sample.} \textbf{a} Schematic render of the proposed PIC for statistical light analysis and control. The layout shows the device under test (DUT), incorporating the unitary layers and the embedded MZIs that drive amplitude modulation. \textbf{b} Optical micrograph of the fabricated sample. False-color blue and yellow overlays highlight the underlying optical waveguide network and the top-level electrical routing traces, respectively. \textbf{c-d} Scanning electron microscopy (SEM) images of the regions highlighted in \textbf{b}. High-resolution details are shown for the five-port waveguide array stage (\textbf{c}) and the Y-branch used in the MZI stage (\textbf{d}).}
    \label{fig:F2}
\end{figure}

The device under test (DUT), illustrated in Fig.~\ref{fig:F2}\textbf{a}, features a five-port and low-depth photonic architecture with a minimal number of components. Ref.~\cite{zelaya2024goldilocks} establishes the required number of active components to universally cover the unitary group $U(N)$. Here, we show that universality is not strictly necessary to perform the coherence analysis of statistical light, and the exact diagonalization condition can be relaxed to the approximation $U(\Phi)V^{\dagger}\approx\mathbb{I}$. Thus, the DUT is devised as a multilayer architecture composed of passive photonic components interleaved by active elements. The passive elements comprise five-port waveguide arrays (WGAs), each using identical waveguides with inhomogeneous spacing to resemble the propagation of the Jx lattice~\cite{Wei16,zelaya2024goldilocks}. 
For prototyping, we employ thermo-optic phase shifters as the active components, allowing for broad implementation in open-access foundries. 

The capabilities of the proposed PIC are further enhanced by incorporating an active layer comprising five Mach-Zehnder interferometers (MZIs) to facilitate amplitude control. Each MZI is constructed with 1x2 Y-branches that split and recombine light. A phase shifter is positioned between the Y-branches to enable amplitude control through destructive interference, while an additional phase shifter at the output of the combiner Y-branch allows for independent phase manipulation. Although this layer breaks network unitarity, it is engineered to function as a pass-through layer when the microheaters between the Y-branches are deactivated.

The PIC was fabricated via an open-access commercial foundry, leveraging a standard fully etched silicon-on-insulator (SOI) platform. The architecture utilizes single-mode silicon waveguides with a uniform cross-section of 500 nm × 220 nm, residing on a 2 µm-thick thermal SiO2 buried oxide layer. A 2.2 µm-thick SiO2 top cladding encapsulates the photonic structures, upon which two discrete metallization layers are deposited. The first layer consists of a high-resistance Ti/W alloy to form highly efficient thermo-optic microheaters, while the second serves as the electrical routing layer.

Fig.~\ref{fig:F2}\textbf{b} shows the optical micrograph of the chip; the underlying waveguide network is visible in bright contrast, overlaid by the yellow contrast of the electrical routing traces. High-resolution scanning electron microscopy (SEM) provides further detail on the precise geometries of the fabrication, specifically highlighting the WGAs (Fig.~\ref{fig:F2}\textbf{c}) and the Y-branches in the MZI stages (Fig.~\ref{fig:F2}\textbf{d}). To ensure stringent modal confinement and suppress evanescent coupling, the WGAs interface with the surrounding circuit components via optimized cubic B\'ezier bends. This geometry fundamentally minimizes both bend-induced propagation losses and crosstalk between adjacent waveguides. Subsequently, 10 µm-radius circular bends are employed to organically fan out the WGA output toward the MZI layer. Operating as proof-of-concept integrated amplitude modulators, each MZI is constructed with compact Y-branches~\cite{zhang2013compact} to govern signal splitting and recombination. While each MZI arm is equipped with a thermo-optic phase shifter, asymmetric driving, actuating only a single arm, proved sufficient to achieve the targeted amplitude modulation depth.

For comprehensive experimental validation, the PIC was fully packaged to enable automated, high-throughput electro-optic characterization. The routing traces terminate in probing pads wire-bonded to a printed circuit board (PCB), providing a robust electrical interface. This permits independent, real-time programming of the thermo-optic microheaters via a computer-controlled external source measurement unit (SMU). Finally, optical input/output is enabled by a pair of 8-degree-angled V-groove fiber arrays with a standard 127 µm pitch, actively aligned and mounted on the chip output grating couplers to ensure stable coupling to the external measurement apparatus.

\subsubsection*{Device Operation, Training, and Assessment of Input Light}
\begin{figure}
    \centering
    \includegraphics[width=0.95\linewidth]{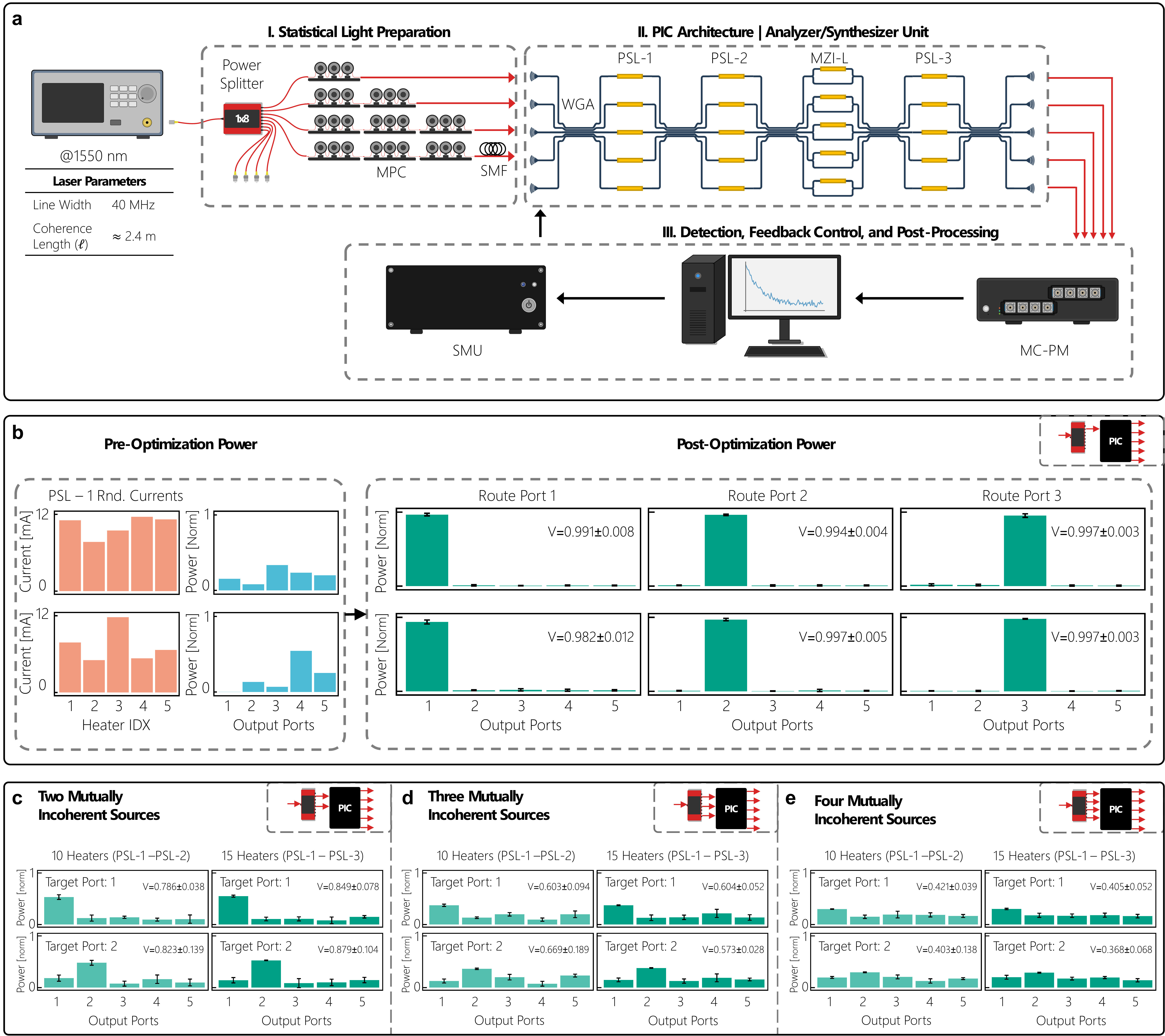}
    \caption{ \textbf{Experimental Setup and Light Source Testing.} 
    \textbf{a} Experimental setup used to perform the coherence analysis and control. This includes a tunable continuous-wave laser (CWL) configured to operate with a broad linewidth, reducing the coherence length. Light is split 1x8 by an optical splitter and routed to the device under test (DUT) via mechanical polarization controllers (MPCs). The length of the MPCs is sufficient to disrupt the CWL coherence length, resulting in three mutually incoherent light sources. An additional incoherent source is generated by attaching a single-mode fiber (SMF). Intensities are recorded using a multi-port power meter (MP-PM), and the microheaters in the DUT are individually driven by a source-measure unit (SMU).
    \textbf{b} Experimental test of fringe generation and visibility characterization when the DUT is supplied with a single source (fully coherent source).
    \textbf{c}-\textbf{e} Visibility assessment for two (\textbf{c}), three (\textbf{d}), and four (\textbf{e}) mutually incoherent sources. For verification purposes, the tests were performed using 10 and 15 heaters. }
    \label{fig:F3}
\end{figure}
The device is characterized and operated using the setup illustrated in Fig.~\ref{fig:F3}\textbf{a}. We employ a tunable continuous-wave laser (CWL, Santec) set to an operational wavelength of 1550 nm. This CWL allows the linewidth to be configured in either a narrow (10 kHz) or broad (40 MHz) regime. To generate mutually incoherent sources, we operate the laser in the broad-linewidth regime, which yields a coherence length of $\ell \approx 2.4$ m. Consequently, breaking the optical coherence requires the optical channels to have path-length differences exceeding $\ell$. To achieve this, we first divide the input signal using a $1\times8$ optical splitter. Each output is routed to a mechanical polarization controller (MPC) that serves a dual purpose: ensuring that the correct transverse-electric (TE) polarization is injected into the PIC and introducing an additional propagation length to break the spatial coherence of the input source. The MPCs used in this experiment comprise a total fiber length of $L_{\textnormal{MPC}} \approx 5.0$ m (satisfying $L_{\textnormal{MPC}} > \ell$). Stacking multiple MPCs thus provides a robust mechanism for preparing statistical light with tunable degrees of coherence. The dashed rectangle in the upper-left panel of Fig.~\ref{fig:F1}\textbf{a} details the configuration for creating mutually incoherent sources across up to four ports. Notably, the optical path differences in the three uppermost contiguous ports are defined by $L_{\textnormal{MPC}}$, which inherently exceeds the coherence length of the CWL. To ensure incoherence at the fourth port, we attach an extra single-mode fiber (SMF) of length $L_{\textnormal{SMF}} \approx 3.0$ m, which also strictly exceeds $\ell$.

The DUT, shown in the upper-right panel of Fig.~\ref{fig:F3}\textbf{a}, showcases three layers of phase shifters (PSL-1, PSL-2, and PSL-3) that induce unitary operations on the input statistical light, and a layer of MZIs (MZI-L). Throughout this section, the heaters on the layer MZI-L are not powered, running the DUT as a unitary device. The DUT is supplied with the prepared incoherent light by selectively connecting the required number of mutually incoherent sources to the fiber array. Output intensity measurements from the PIC are recorded using a multichannel power meter (MC-PM) capable of simultaneous multi-port sampling. A control computer records these power measurements and interfaces the source measurement unit (SMU), an electronic driver that individually powers each microheater on the DUT. This setup establishes an automated framework for dynamically updating the applied current in each heater while synchronously recording the corresponding optical power at the DUT output. Due to the large number of microheaters, we operate the PIC with an external thermoelectric cooler (TEC) to maintain the chip temperature at 27 ${}^{\circ}$C.

Before proceeding, we conduct a preliminary test to verify the incoherence of the supplied light. A defining characteristic of mutually incoherent light is the absence of interference fringes. Traditionally, this verification is performed off-chip in the far field by interfering the sources pairwise via a beam-splitter and projecting the resulting pattern onto a screen. In this work, we probe for such interference directly on-chip by appropriately programming the DUT. To quantify this, we define the unit-vector figure of merit
$$\mathcal{L}(\Phi;n) = \frac{\Vert \mathbf{P}(\Phi) - \hat{e}_{n} \Vert^2}{N}, 
\label{eq:FoM3}$$
where $\hat{e}_{n}\in\mathbb{R}^{N}$ is a unit vector with a value of one at the $n$-th position and zeros elsewhere, and $\mathbf{P}(\Phi)$ is a normalized vector containing the measured optical powers.

For a fully coherent source, the target unit vector can always be achieved by routing all optical power into a single output port. Although the proposed five-port DUT does not render a universal unitary device, previous work has demonstrated that only two layers are required to generate arbitrary spatial photonic states~\cite{zelaya2025chip}, including the desired unit-vector states. Consequently, we first evaluate the ability of the DUT to vanish optical power at specific output ports by exciting it with a single and fully coherent source. To ensure the optimization is robust and not merely accidental, we initialize the PSL-1 layer with random drive currents sampled uniformly from the interval $(0, 12.5)$ mA. Subsequently, both the PSL-1 and PSL-2 layers are actively tuned during the optimization process to minimize Eq.~\eqref{eq:FoM3}.

Fig.~\ref{fig:F3}\textbf{b} illustrates the results of multiple optimization trials initiated from two distinct sets of random starting currents. The left panel displays the initial random current distributions applied to PSL-1 (red bars) alongside the corresponding unoptimized power distributions at the DUT output (blue bars). Following multiple optimization cycles for each random initialization, the mean and standard deviation of the final power distributions are plotted in the right panel of Fig.~\ref{fig:F3}\textbf{b} for target unit vectors $n=1, 2,$ and $3$. These results clearly demonstrate that the DUT can reliably route power to a single target port—producing high-contrast output distributions if the supplied source is coherent. This performance is formally quantified via the interference visibility~\cite{mandel1995optical}
\begin{equation*}
    V=\frac{\textnormal{max}(\mathbf{P})-\textnormal{min}(\mathbf{P})}{\textnormal{max}(\mathbf{P}) + \textnormal{min}(\mathbf{P})}, 
\label{eq:vis}
\end{equation*}
which yields $V=1$ for fully coherent sources and $V=0$ for mutually incoherent sources, with intermediate values indicating partial coherence. As shown in the right panel of Fig.~\ref{fig:F3}\textbf{b}, the experimental results exhibit excellent agreement with the expected visibility.

We repeat this test using two, three, and four mutually incoherent sources, executing multiple optimizations for each configuration. The corresponding results, summarized in Figs.~\ref{fig:F3}\textbf{c}--\textbf{e}, reveal that these configurations are unable to generate the required zeros and sharp peaks for target unit vectors $n=0, 1$. Similar behavior is observed across all other target unit vectors. To determine whether this inability to suppress optical power stems from an insufficient number of active phase elements, we conduct the optimization in two separate sequences: the first utilizes 10 active heaters (comprising PSL-1 and PSL-2), while the second employs 15 heaters (spanning PSL-1 through PSL-3). The experimental results indicate that the interference visibility remains approximately constant for a given incoherent source, independent of the number of heaters engaged during the optimization. Furthermore, we observe a consistent, monotonic decrease in visibility as the number of input sources increases, confirming their mutual incoherence.

\subsubsection*{Analysis and Control of Statistical Light}

\begin{figure}
    \centering
    \includegraphics[width=0.95\linewidth]{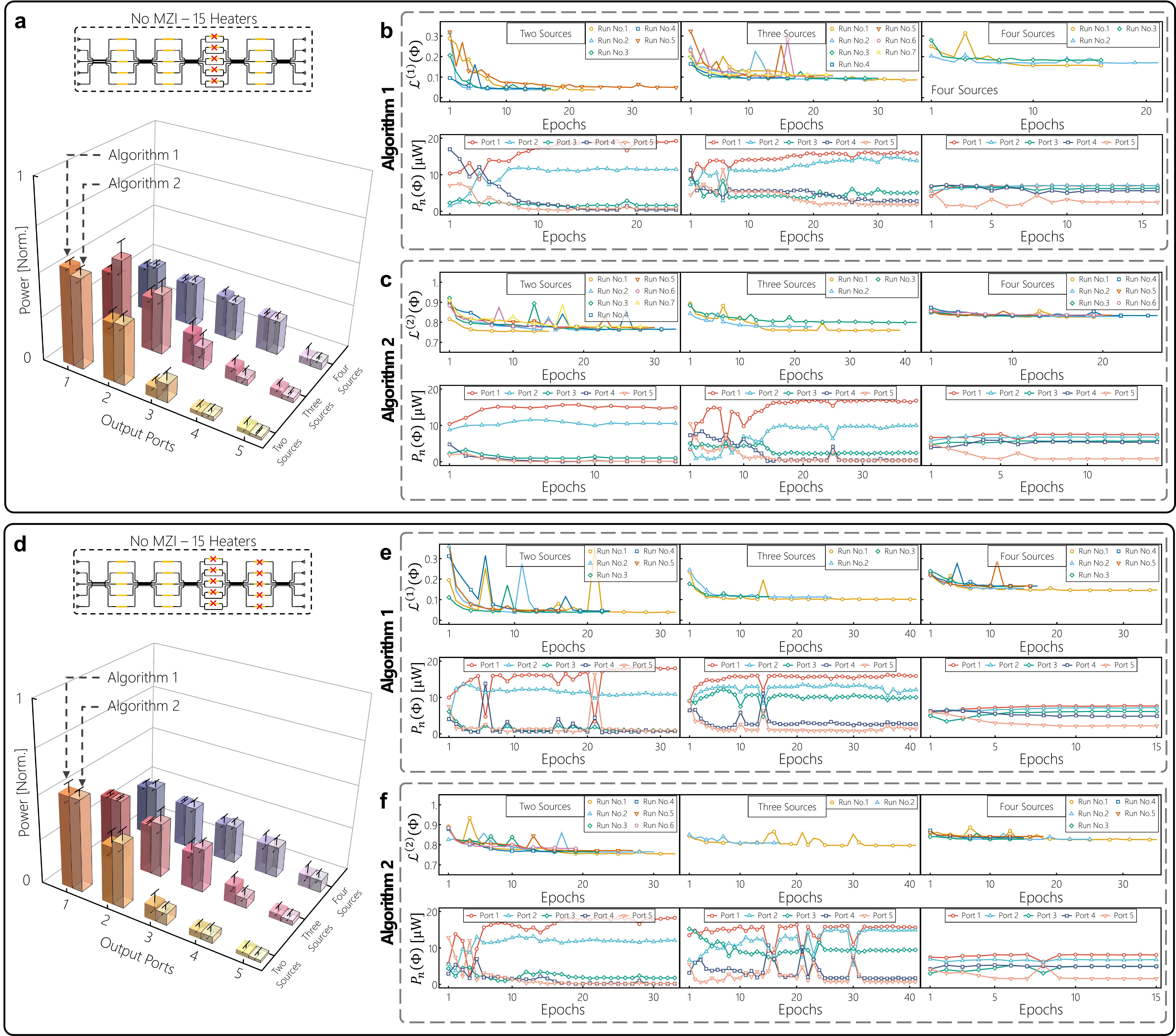}
    \caption{\textbf{Experimental reconstruction of coherence matrix eigenvalues via architecture-agnostic optimization.} 
\textbf{a}–\textbf{c} Eigenvalue extraction utilizing a 15-heater configuration (three phase-shifter layers). Reconstructed eigenvalues (\textbf{a}) achieved by minimizing the cumulative sum ($\mathcal{L}^{(1)}$) and product ($\mathcal{L}^{(2)}$) FoMs across multiple runs. Top panels in (\textbf{b}) and (\textbf{c}) display the training dynamics for $\mathcal{L}^{(1)}$ and $\mathcal{L}^{(2)}$, respectively, when the device is illuminated by two (left), three (middle), and four (right) mutually incoherent sources. Bottom panels track the corresponding temporal evolution of the normalized output port powers for the optimal runs. The observed suppression of output expressivity for four sources physically captures the transition toward a fully incoherent regime.
\textbf{d}–\textbf{f} Resource-efficient eigenvalue extraction utilizing a shallower 10-heater configuration (two phase-shifter layers). Following an identical experimental and filtering protocol as the 15-heater setup, these results demonstrate that reduced hardware depth retains sufficient degrees of freedom to accurately resolve the hidden eigenvalues, validating the scalability of the protocol. }
    \label{fig:F4}
\end{figure}

\textbf{Coherence Analysis.} Although visibility tests offer a preliminary assessment of input coherence, a rigorous quantification demands the extraction of the coherence matrix eigenvalues. Although our experimental synthesis of mutually incoherent sources provides a priori knowledge of the expected eigenvalue distributions, our architecture-agnostic approach and proposed figures of merit (FoMs) extract these fundamental properties without assumptions about the input signal. To demonstrate this capability, we evaluated two distinct operational regimes: a 15-heater configuration (spanning phase-shifter layers PSL-1 through PSL-3) and a shallower 10-heater configuration (PSL-1 and PSL-2), with the MZI-L heaters remaining inactive.

The experimental eigenvalue reconstruction for the 15-heater configuration is detailed in Figs.~\ref{fig:F4}\textbf{a}--\textbf{c}. Figure~\ref{fig:F4}\textbf{a} maps the reconstructed eigenvalues obtained by optimizing the cumulative sum and product FoMs [Eq.~\eqref{eq:loss1}] across multiple randomized initial states. In physical implementations, hardware noise and local minima can occasionally drive the algorithm toward unsorted normalized output powers. To rigorously mitigate this, we implement a post-selection filter with a predefined tolerance threshold ($\delta_{th}=0.02$) to accommodate nearly degenerate eigenvalues. We retain only the physical solutions that satisfy the ordering condition
\begin{equation*}
    \textnormal{max}\left(\{P_{n+1}-P_{n}\}_{n=1}^{N-1}\right)<\delta_{th},
\end{equation*}
discarding any anomalous convergence pathways.

The training dynamics for the cumulative sum ($\mathcal{L}^{(1)}$) and product ($\mathcal{L}^{(2)}$) FoMs are illustrated in the top panels of Figs.~\ref{fig:F4}\textbf{b} and \textbf{c}, respectively, across inputs comprising two, three, and four mutually incoherent sources. The non-zero convergence bounds of these FoMs naturally arise from inverting the maximization problem into a minimization framework, as formalized in Eq.~\eqref{eq:loss1}. Critically, the lower panels track the evolution of the output port power for the optimal runs. These trajectories reveal a profound physical consequence: feeding the device with four mutually incoherent sources severely suppresses the expressivity of the output optical power. This scaling captures the transition into a fully incoherent regime, in which both expressivity and visibility vanish.

To demonstrate the resource efficiency of our protocol, we extended the eigenvalue extraction to the 10-heater configuration. This is supported by numerical predictions indicating that two phase-shifter layers provide sufficient degrees of freedom to accurately reconstruct the hidden eigenvalues. Following an identical experimental protocol, we evaluated both algorithms to enable a direct performance comparison across hardware depths (Figs.~\ref{fig:F4}\textbf{d}-\textbf{f}). For clarity, Table~\ref{tab:source_configurations} presents the experimental results, demonstrating that the reconstructed eigenvalues are consistently robust across all experimental runs, algorithms, and source types.\\

\begin{table}[t]
\centering
\caption{\textbf{Summarized results.} Reconstructed hidden eigenvalues using the configuration with 15 heaters (15H) and 10 heaters (10H), in combination with the cumulative sum (A1) and cumulative product (A2) FoMs, for different numbers of mutually incoherent sources.}
\label{tab:source_configurations}
\begin{tabular*}{\textwidth}{@{\extracolsep{\fill}} lll ccccc @{}}
\toprule
\multicolumn{3}{l}{\multirow{2}{*}{Configuration}} & \multicolumn{5}{c}{Output Ports} \\ 
\cmidrule{4-8} 
\multicolumn{3}{c}{}                               & 1 & 2 & 3 & 4 & 5 \\ 
\midrule
\multirow{4}{*}{\begin{tabular}[c]{@{}l@{}}Two \\ Sources\end{tabular}}   & \multirow{2}{*}{15H} & A1 & $0.55\pm 0.02$ & $0.34\pm 0.05$ & $0.06\pm 0.03$ & $0.03\pm 0.01$ & $0.03\pm 0.02$ \\
                               &                      & A2 & $0.52\pm 0.03$ & $0.34\pm 0.04$ & $0.10\pm 0.05$ & $0.02\pm 0.02$ & $0.02\pm 0.01$ \\ \addlinespace
                               & \multirow{2}{*}{10H} & A1 & $0.53\pm 0.05$ & $0.33\pm 0.06$ & $0.08\pm 0.05$ & $0.04\pm 0.02$ & $0.02\pm 0.01$ \\
                               &                      & A2 & $0.52\pm 0.03$ & $0.36\pm 0.05$ & $0.07\pm 0.02$ & $0.04\pm 0.03$ & $0.01\pm 0.01$ \\ 
\midrule
\multirow{4}{*}{\begin{tabular}[c]{@{}l@{}}Three \\ Sources\end{tabular}} & \multirow{2}{*}{15H} & A1 & $0.38\pm 0.02$ & $0.31\pm 0.04$ & $0.19\pm 0.05$ & $0.08\pm 0.03$ & $0.04\pm 0.02$ \\
                               &                      & A2 & $0.47\pm 0.09$ & $0.35\pm 0.06$ & $0.12\pm 0.07$ & $0.04\pm 0.03$ & $0.03\pm 0.01$ \\ \addlinespace
                               & \multirow{2}{*}{10H} & A1 & $0.37\pm 0.01$ & $0.28\pm 0.01$ & $0.21\pm 0.04$ & $0.10\pm 0.04$ & $0.04\pm 0.02$ \\
                               &                      & A2 & $0.36\pm 0.01$ & $0.31\pm 0.06$ & $0.25\pm 0.03$ & $0.05\pm 0.03$ & $0.03\pm 0.01$ \\ 
\midrule
\multirow{4}{*}{\begin{tabular}[c]{@{}l@{}}Four \\ Sources\end{tabular}}   & \multirow{2}{*}{15H} & A1 & $0.26\pm 0.02$ & $0.25\pm 0.01$ & $0.24\pm 0.02$ & $0.20\pm 0.01$ & $0.04\pm 0.05$ \\
                               &                      & A2 & $0.28\pm 0.01$ & $0.26\pm 0.01$ & $0.22\pm 0.01$ & $0.20\pm 0.02$ & $0.04\pm 0.01$ \\ \addlinespace
                               & \multirow{2}{*}{10H} & A1 & $0.28\pm 0.02$ & $0.26\pm 0.01$ & $0.21\pm 0.03$ & $0.18\pm 0.04$ & $0.07\pm 0.04$ \\
                               &                      & A2 & $0.29\pm 0.02$ & $0.25\pm 0.03$ & $0.21\pm 0.02$ & $0.18\pm 0.04$ & $0.07\pm 0.03$ \\ 
\midrule
\bottomrule
\end{tabular*}
\end{table}


\textbf{Coherence Control.} As dictated by the eigendecomposition of the coherence matrix, unitary operations inherently conserve the statistical nature of light. Modifying this statistical state requires breaking the unitary regime (detailed in the Materials section). The MZI layer embedded within the PIC provides this critical functionality by introducing non-unitary operations that enhance coherence metrics such as visibility. Although the shallow depth of our current DUT is well-suited for coherence analysis, statistical light control ultimately demands additional active elements. Nevertheless, to demonstrate a foundational coherence control protocol, we inject two mutually incoherent sources into the DUT, a configuration that initially yields a visibility of $V\sim 0.84$ (Fig.~\ref{fig:F3}\textbf{b}).

To actively tailor this coherence, we apply the unit-vector optimization defined in Eq.~\eqref{eq:FoM3}, expanding the parameter space $\Phi$ to encompass the phase shifters within MZI-L. To systematically monitor device losses, we record the baseline optical power prior to initializing the randomized currents. We then perform multiple optimization trials to route the light independently to output ports $n=1, 2,$ and $3$, selecting the state that minimizes the FoM for each routing target.

\begin{figure}[t]
    \centering
    \includegraphics[width=0.95\linewidth]{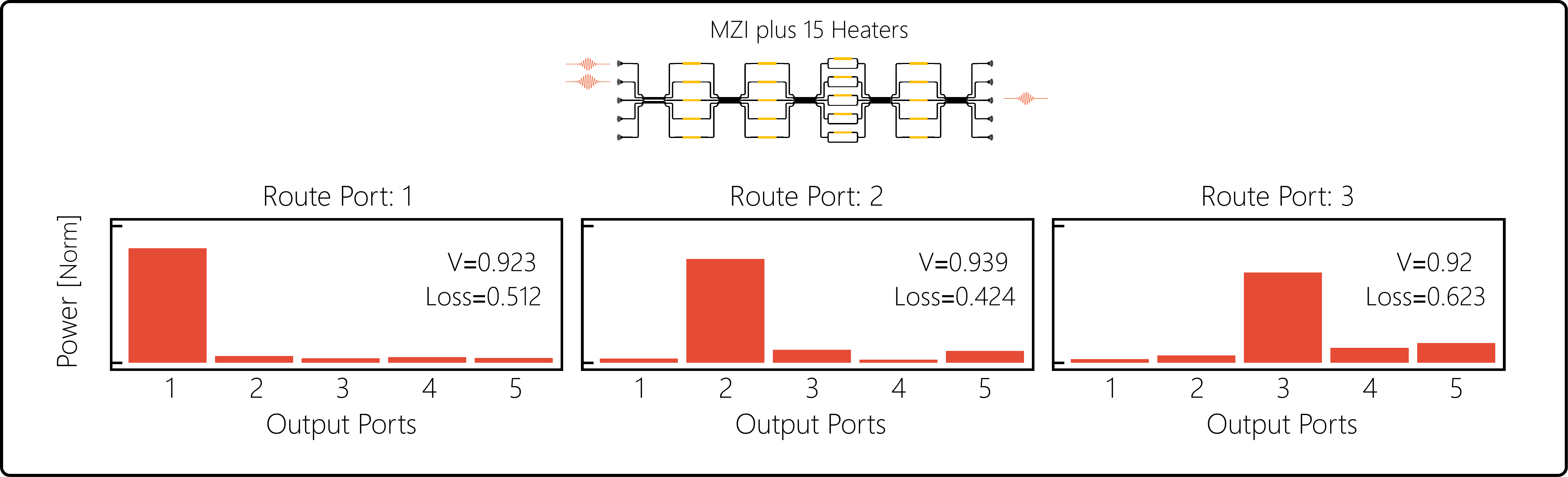}
    \caption{\textbf{Statistical Light Control.} Experimental configuration of the DUT for active coherence manipulation via tunable amplitude modulation within the MZI layer. The control protocol is demonstrated by injecting two mutually incoherent sources at the input. The resulting coherence enhancement is quantified by measuring the final visibility, obtained by executing the unit-vector optimization [Eq.~\eqref{eq:FoM3}] with the MZI heaters actively included in the optimization parameter space.}
    \label{fig:F5}
\end{figure}

The resulting coherence-controlled states (Fig.~\ref{fig:F5}) demonstrate a substantial enhancement in visibility, reaching $ V \sim 0.93$. However, this protocol exposes a fundamental physical trade-off. As shown in Fig.~\ref{fig:F5}, improving visibility inevitably incurs a loss penalty due to the requisite amplitude modulation induced across the MZIs.

\section*{Discussion} 

We have introduced an architecture-agnostic, in-situ method for analyzing the spatial optical coherence of radiation fields using programmable PICs. By leveraging the Schur-Horn theorem, our framework treats the unitary network as a black box, effectively circumventing the need for prior knowledge of the internal network topology or the incident light. By minimizing specialized cumulative sum and product FoMs via gradient-descent optimization, we successfully demonstrated the direct extraction of the coherence matrix eigenvalues using only output-intensity measurements.  

Our experimental findings validate the robustness and resource efficiency of this approach. During physical implementation, we observed that hardware noise and local minima can occasionally drive the algorithm toward unsorted output powers. However, applying a post-selection filter with a predefined tolerance threshold ($\delta_{th}=0.02$) successfully discarded these anomalous pathways, ensuring that the eigenvalue distributions were strictly sorted and physically valid. Notably, achieving accurate coherence extraction does not require a deep, universally parameterized architecture. By executing the optimization protocol on a shallower, 10-heater configuration, we demonstrated that reduced hardware depth retains sufficient degrees of freedom to reconstruct the hidden eigenvalues. This resource efficiency highlights the viability of deploying coherence analysis on ultra-compact devices with a reduced physical footprint.  

Tracking the experimental training dynamics provided fundamental physical insights into the behavior of statistical light within the PIC. As we incrementally fed the device with two, three, and four mutually incoherent sources, the resulting optical expressivity was significantly suppressed. This monotonic suppression perfectly captures the physical transition toward a fully incoherent regime, in which both interference visibility and output expressivity vanish.  

Beyond purely diagnostic analysis, we demonstrated the capability to actively tailor the coherence properties of light on-chip. Because unitary transformations inherently conserve the statistical nature of light, modifying this state requires breaking from the unitary regime. By activating the embedded MZI layer, we introduced non-unitary amplitude modulation. This foundational coherence-control protocol successfully increased the interference visibility of two mutually incoherent sources from an initial value of $V \sim 0.84$ to $V \sim 0.93$. Nevertheless, our findings reveal a fundamental physical trade-off; that is, improving spatial coherence via this mechanism inevitably incurs an optical loss penalty due to the destructive interference required for amplitude modulation.  

Ultimately, this architecture-agnostic framework proves highly adaptable. Because the operations are inherently programmable, the unitary network can be explicitly configured to yield the identity matrix upon completion of the eigenvalue extraction, allowing the analyzed field to pass through the device unaffected for subsequent downstream applications. By establishing a robust mechanism that operates effectively even under lossy, non-universal conditions, this work provides a highly scalable paradigm for realizing ultra-compact spatial coherence analyzers and programmable statistical light controllers.


\section*{Methods}

\textbf{Coherence Matrix Properties.} Let us assume that the partially coherent field sampled at the $N$ points is fed into a general linear transform $A\in GL(N)$. Following the definition of the coherence matrix, and assuming that the changes in time of the linear transformer are much larger than the statistical fluctuations of the fields, one can see that the coherence matrix produced at the output of the linear transformer becomes 
\begin{equation}
\rho_{\textnormal{out}}= \langle \mathbf{x}_{\textnormal{out}}\mathbf{x}_{\textnormal{out}}^{\dagger} \rangle = A\rho A^{\dagger} , \quad \mathbf{x}_{\textnormal{out}}:=A\mathbf{x} .
\label{eq:field2}
\end{equation} 
The degree of coherence is measured by the rank of the coherence matrix; however, reconstructing the full coherence matrix is challenging and, in some experimental setups, even prohibitive. In turn, it is more convenient to reconstruct only its eigenvalues, which already encode the degree of coherence and can be used to analyze the coherence content, as proven successful in Ref.~\cite{roques2024measuring}. 

The structure of the coherence matrix can be modified via two mechanisms. First, via unitary control, $A\in U(N)$, which shuffles the correlation components while preserving the degree of coherence. This is a handy resource for controlling the scattering and absorption dynamics of radiation fields before reaching the scatterer~\cite{guo2024unitaryA,guo2024unitaryB}. Secondly, by introducing losses into the network, which modifies the eigenvalues of the coherence matrix and thus alters the degree of coherence. These losses break the unitary evolution and induce a general linear transform $A\in GL(N)\supset U(N)$ instead. Still, the output density matrix transforms as $\rho_{\textnormal{out}}=A\rho_{\textnormal{in}}A^{\dagger}$ and satisfies the coherence matrix conditions.

The diagonal component $\rho_{n,n}$ is proportional to the intensities at the $n$-th sampling point, for $n\in\{1,\ldots,N\}$. The total intensity is proportional to the trace of $\rho$, which, without loss of generality, we normalize to unity ($\textnormal{tr}\rho=1$). We begin the analysis by focusing on parameterized $N$-dimensional unitary matrices, $U(\Phi):\mathbb{R}^{K}\rightarrow \mathbb{C}^{N\times N}$, which are not necessarily universal. Here, $K$ denotes the total number of real-valued parameters in $\Phi=\{\phi_{k}\}_{k=1}^{K}$ controlling the transfer matrix of the unitary network, the specifics of which depend on the network design.\\

\textbf{Schur-Horn Theorem for Coherence Analysis.} Since the $N$ power measurements at the network output sample the diagonal components of the output coherence matrix, and the network performs unitary operations, the eigenvalues of the output coherence matrix $\rho_{\textnormal{out}}$ are preserved. Thus, to extract the coherence matrix eigenvalues, we shall ensure that the output coherence matrix is diagonal; in that case, the power measurements correspond, up to a normalization factor, to the coherence matrix eigenvalues. 

Although full characterization of the coherence matrix is beyond the scope of this letter, the diagonalization condition can be assessed only by measuring the output power. This can be ensured via the Schur–Horn theorem~\cite{horn1954doubly}, which summarizes as follows: let $X=(x_{1},\ldots, x_{N})\in\mathbb{R}^{N}$ and $Y=(y_{1},\ldots, y_{N})\in\mathbb{R}^{N}$ be two sequences of non-increasing real numbers; then, there is a Hermitian matrix $H$ with diagonal components $Y$ and eigenvalues $X$ if and only if {\small $\sum_{n=1}^{K}(x_{n}-y_{n})\geq 0$}, for all $K=\{1,\ldots,N\}$. Since coherence matrices satisfy this condition, normalized power measurements will never exceed the normalized coherence matrix eigenvalues. Furthermore, the equality in Schur-Horn's theorem holds when the coherence matrix is already diagonal. \\

\noindent\textbf{Funding.} This project is supported by the U.S. Air Force Office of Scientific Research (AFOSR) Award\# FA9550-25-1-0200. \\

\noindent\textbf{Disclosures.} The authors declare no conflicts of interest.\\

\noindent\textbf{Data availability.} Data underlying the results presented in this paper can be obtained from the authors upon reasonable request. 


\bibliography{biblio}

@book{goodman2015statistical,
  title={Statistical Optics},
  author={Goodman, Joseph W},
  year={2015},
  publisher={John Wiley \& Sons},
  address   = {New Jersey},
  isbn      = {978-1-119-00945-0}
}

@article{wolf1982new,
  title={New theory of partial coherence in the space--frequency domain. Part I: spectra and cross spectra of steady-state sources},
  author={Wolf, Emil},
  journal={Journal of the Optical Society of America},
  volume={72},
  number={3},
  pages={343--351},
  year={1982},
  publisher={Optica Publishing Group}
}

@article{Wei16,
author={Weimann, Steffen and Perez-Leija, Armando and Lebugle, Maxime and Keil, Robert and Tichy, Malte and Grafe, Markus and Heilmann, Ren{\'e} and Nolte, Stefan and Moya-Cessa, Hector and Weihs, Gregor and Christodoulides, Demetrios N. and Szameit, Alexander},
title={Implementation of quantum and classical discrete fractional Fourier transforms},
journal={Nature Communications},
year={2016},
month={Mar},
day={23},
volume={7},
number={1},
pages={11027},
issn={2041-1723},
doi={10.1038/ncomms11027},
}

@article{friedman2025programmable,
  title={Programmable space-frequency linear transformations in photonic interlacing architectures},
  author={Friedman, Jonathan and others},
  journal={Scientific Reports},
  volume={15},
  number={1},
  pages={35173},
  year={2025},
  publisher={Nature Publishing Group UK London}
}

@article{de2018simple,
  title={Simple factorization of unitary transformations},
  author={de Guise, Hubert and Di Matteo, Olivia and S{\'a}nchez-Soto, Luis L},
  journal={Physical Review A},
  volume={97},
  number={2},
  pages={022328},
  year={2018},
  publisher={APS}
}

@article{zhang2013compact,
  title={A compact and low loss Y-junction for submicron silicon waveguide},
  author={Zhang, Yi and Yang, Shuyu and Lim, Andy Eu-Jin and Lo, Guo-Qiang and Galland, Christophe and Baehr-Jones, Tom and Hochberg, Michael},
  journal={Optics express},
  volume={21},
  number={1},
  pages={1310--1316},
  year={2013},
  publisher={Optica Publishing Group}
}

@misc{markowitz2023universal,
  title={Universal unitary photonic circuits by interlacing discrete fractional Fourier transform and phase modulation},
  author={Markowitz, Matthew and Miri, Mohammad-Ali},
  note={arXiv:2307.07101 [physics.optics], 2023}
}

@article{Huang94,
  title={Coupled-mode theory for optical waveguides: an overview},
  author={Huang, Wei-Ping},
  journal={JOSA A},
  volume={11},
  number={3},
  pages={963--983},
  year={1994},
  publisher={Optica Publishing Group},
  doi = {10.1364/JOSAA.11.000963}
}

@article{zelaya2025chip,
  title={On-chip unitary generation of arbitrary complex spatial photonic states},
  author={Zelaya, Kevin and Honari-Latifpour, Mostafa and Mandal, Kishor K and Friedman, Jonathan and Madamopoulos, Nicholas and Miri, Mohammad-Ali},
  journal={Optica},
  volume={12},
  number={9},
  pages={1492--1501},
  year={2025},
  publisher={Optica Publishing Group}
}

@article{bogaerts2020programmable,
  title={Programmable photonic circuits},
  author={Bogaerts, Wim and P{\'e}rez, Daniel and Capmany, Jos{\'e} and Miller, David AB and Poon, Joyce and Englund, Dirk and Morichetti, Francesco and Melloni, Andrea},
  journal={Nature},
  volume={586},
  number={7828},
  pages={207--216},
  year={2020},
  publisher={Nature Publishing Group UK London}
}

@article{reck1994experimental,
  title={Experimental realization of any discrete unitary operator},
  author={Reck, Michael and Zeilinger, Anton and Bernstein, Herbert J and Bertani, Philip},
  journal={Physical review letters},
  volume={73},
  number={1},
  pages={58},
  year={1994},
  publisher={APS}
}

@article{clements2016optimal,
  title={Optimal design for universal multiport interferometers},
  author={Clements, William R and Humphreys, Peter C and Metcalf, Benjamin J and Kolthammer, W Steven and Walmsley, Ian A},
  journal={Optica},
  volume={3},
  number={12},
  pages={1460--1465},
  year={2016},
  publisher={Optical Society of America},
  doi={10.1364/OPTICA.3.001460}
}

@article{Shokraneh2020,
  title={The diamond mesh, a phase-error-and loss-tolerant field-programmable MZI-based optical processor for optical neural networks},
  author={Shokraneh, Farhad and Geoffroy-Gagnon, Simon and Liboiron-Ladouceur, Odile},
  journal={Optics Express},
  volume={28},
  number={16},
  pages={23495--23508},
  year={2020},
  publisher={Optica Publishing Group},
  doi={10.1364/OE.395441}
}

@article{rahbardar2023addressing,
  title={Addressing the programming challenges of practical interferometric mesh based optical processors},
  author={Rahbardar Mojaver, Kaveh and Zhao, Bokun and Leung, Edward and Safaee, S and Liboiron-Ladouceur, Odile},
  journal={Optics Express},
  volume={31},
  pages={23851-23866},
  year={2023},
  doi={10.1364/OE.489493}
}

@article{yi2021multi,
  title={Multi-functional photonic processors using coherent network of micro-ring resonators},
  author={Yi, Dan and Wang, Yi and Tsang, Hon Ki},
  journal={APL Photonics},
  volume={6},
  number={10},
  year={2021},
  publisher={AIP Publishing}
}

@article{miller2013self,
  title={Self-configuring universal linear optical component},
  author={Miller, David AB},
  journal={Photonics Research},
  volume={1},
  number={1},
  pages={1--15},
  year={2013},
  publisher={Optica Publishing Group}
}

@article{tanomura2022scalable,
  title={Scalable and robust photonic integrated unitary converter based on multiplane light conversion},
  author={Tanomura, Ryota and Tang, Rui and Umezaki, Toshikazu and Soma, Go and Tanemura, Takuo and Nakano, Yoshiaki},
  journal={Physical Review Applied},
  volume={17},
  number={2},
  pages={024071},
  year={2022},
  publisher={APS}
}

@article{tang2021ten,
  title={Ten-port unitary optical processor on a silicon photonic chip},
  author={Tang, Rui and Tanomura, Ryota and Tanemura, Takuo and Nakano, Yoshiaki},
  journal={Acs Photonics},
  volume={8},
  number={7},
  pages={2074--2080},
  year={2021},
  publisher={ACS Publications}
}

@article{pastor2021arbitrary,
  title={Arbitrary optical wave evolution with {F}ourier transforms and phase masks},
  author={Pastor, V{\'\i}ctor L{\'o}pez and Lundeen, Jeff and Marquardt, Florian},
  journal={Optics Express},
  volume={29},
  number={23},
  pages={38441--38450},
  year={2021},
  publisher={Optical Society of America}
}

@article{Markowitz2023auto,
  title={Auto-calibrating universal programmable photonic circuits: hardware error-correction and defect resilience},
  author={Markowitz, Matthew and Zelaya, Kevin and Miri, Mohammad-Ali},
  journal={Optics Express},
  volume={31},
  number={23},
  pages={37673--37682},
  year={2023},
  publisher={Optica Publishing Group},
  doi={10.1364/OE.502226}
}

@article{zelaya2024goldilocks,
  title={The Goldilocks principle of learning unitaries by interlacing fixed operators with programmable phase shifters on a photonic chip},
  author={Zelaya, Kevin and Markowitz, Matthew and Miri, Mohammad-Ali},
  journal={Scientific Reports},
  volume={14},
  number={1},
  pages={10950},
  year={2024},
  publisher={Nature Publishing Group UK London},
  doi = {10.1038/s41598-024-60700-8}
}

@article{saygin_robust_2020,
	title = {Robust {Architecture} for {Programmable} {Universal} {Unitaries}},
	volume = {124},
	doi = {10.1103/PhysRevLett.124.010501},
	number = {1},
	urldate = {2022-10-24},
	journal = {Physical Review Letters},
	author = {Saygin, M. Yu. and Kondratyev, I. V. and Dyakonov, I. V. and Mironov, S. A. and Straupe, S. S. and Kulik, S. P.},
	year = {2020},
	pages = {010501}
}

@book{born2013principles,
  title={Principles of optics: electromagnetic theory of propagation, interference and diffraction of light},
  edition= {7th expanded},
  author={Born, Max and Wolf, Emil},
  year={2013},
  publisher={Cambridge University Press},
  address={Cambridge},
  doi= {10.1017/CBO9781139644181}
}

@article{gabor1948new,
  title={A new microscopic principle},
  author={Gabor, Dennis},
  journal={Nature},
  volume={161},
  number={4098},
  pages={777--778},
  year={1948},
  publisher={Nature Publishing Group UK London}
}

@article{huang2024quantitative,
  title={Quantitative phase imaging based on holography: trends and new perspectives},
  author={Huang, Zhengzhong and Cao, Liangcai},
  journal={Light: Science \& Applications},
  volume={13},
  number={1},
  pages={145},
  year={2024},
  publisher={Nature Publishing Group UK London}
}

@article{huang1991optical,
  title={Optical coherence tomography},
  author={Huang, David and Swanson, Eric A and Lin, Charles P and Schuman, Joel S and Stinson, William G and Chang, Warren and Hee, Michael R and Flotte, Thomas and Gregory, Kenton and Puliafito, Carmen A and Fujimoto, James G},
  journal={Science},
  volume={254},
  number={5035},
  pages={1178--1181},
  year={1991},
  publisher={American Association for the Advancement of Science}
}

@book{goodman2007speckle,
title={Speckle Phenomena in Optics: Theory and Applications},
author={Goodman, Joseph W},
address={Bellingham},
year={2010},
publisher={SPIE Press},
edition={Second},
}

@book{bohman2006optical,
title={Optical Coherence Tomography: Principles and Applications},
author={Bohman, W and others},
address={Cambridge},
year={2006}
}

@book{siegman1986lasers,
title={Lasers},
author={Siegman, Anthony E},
year={1986},
address={Mill Valley},
publisher={University Science Books}
}

@book{attwood2017xrays,
  title={X-Rays and Extreme Ultraviolet Radiation: Principles and Applications},
  author={Attwood, David T and Sakdinawat, Anne},
  year={2017},
  edition={2nd},
  publisher={Cambridge University Press},
  address={Cambridge}
}

@article{mcneil2010xray,
  title={X-ray free-electron lasers},
  author={McNeil, Brian WJ and Thompson, Neil R},
  journal={Nature Photonics},
  volume={4},
  number={12},
  pages={814--821},
  year={2010},
  publisher={Nature Publishing Group},
  doi={doi.org/10.1038/nphoton.2010.239}
}

@article{pellegrini2016physics,
  title={The physics of x-ray free-electron lasers},
  author={Pellegrini, Claudio and Marinelli, Agostino and Reiche, Sven},
  journal={Reviews of Modern Physics},
  volume={88},
  number={1},
  pages={015006},
  year={2016},
  publisher={American Physical Society},
  doi={doi.org/10.1103/RevModPhys.88.015006}
}

@book{mandel1995optical,
  title={Optical Coherence and Quantum Optics},
  author={Mandel, Leonard and Wolf, Emil},
  year={1995},
  publisher={Cambridge University Press},
  address={Cambridge},
  isbn={9780521417112},
  url={https://www.cambridge.org/core/books/optical-coherence-and-quantum-optics/F8CB94C70FA64CD3FB60890CA2048168}
}

@article{roques2024measuring,
  title={Measuring, processing, and generating partially coherent light with self-configuring optics},
  author={Roques-Carmes, Charles and Fan, Shanhui and Miller, David AB},
  journal={Light: Science \& Applications},
  volume={13},
  number={1},
  pages={260},
  year={2024},
  publisher={Nature Publishing Group UK London}
}

@article{guo2024unitaryA,
  title={Unitary control of partially coherent waves. I. Absorption},
  author={Guo, Cheng and Fan, Shanhui},
  journal={Physical Review B},
  volume={110},
  number={3},
  pages={035430},
  year={2024},
  publisher={APS}
}

@article{guo2024unitaryB,
  title={Unitary control of partially coherent waves. II. Transmission or reflection},
  author={Guo, Cheng and Fan, Shanhui},
  journal={Physical Review B},
  volume={110},
  number={3},
  pages={035431},
  year={2024},
  publisher={APS}
}

@article{hashemi2026chip,
  title={On-chip control of the coherence matrix of four-mode partially coherent light: rank, entropy, and modal Stokes parameters},
  author={Hashemi, Amin and Shiri, Abbas and Saleh, Bahaa EA and Blanco-Redondo, Andrea and Abouraddy, Ayman F},
  journal={arXiv:2601.18797},
  year={2026}
}

@article{horn1954doubly,
  title={Doubly stochastic matrices and the diagonal of a rotation matrix},
  author={Horn, Alfred},
  journal={American Journal of Mathematics},
  volume={76},
  number={3},
  pages={620--630},
  year={1954},
  publisher={JSTOR}
}

@article{lee2020light,
  title={Light source optimization for partially coherent holographic displays with consideration of speckle contrast, resolution, and depth of field},
  author={Lee, Seungjae and Kim, Dongyeon and Nam, Seung-Woo and Lee, Byounghyo and Cho, Jaebum and Lee, Byoungho},
  journal={Scientific reports},
  volume={10},
  number={1},
  pages={18832},
  year={2020},
  publisher={Nature Publishing Group UK London}
}

@misc{zelaya2026analysis,
  title={Architecture-agnostic analysis of partially coherent light with programmable photonics},
  author={Kevin Zelaya and Matthew Markowitz and Jonathan Friedman and Mohammad-Ali Miri},
  note={arXiv:2607.24104v1 [physics.optics]}
}

\end{document}